\documentclass[twocolumn,showpacs,aps,prl,superscriptaddress]{revtex4}
\usepackage[dvips]{graphicx}
\usepackage{latexsym}
\def\bslantfrac#1#2{{#1}\backslash\kern-0.1em{#2}}

\begin{document}
\newcommand{\Real}{\Re\text{e}}
\newcommand{\Imag}{\Im\text{m}}


\title{Deeply Virtual Compton Scattering Beam-Spin Asymmetries}


\newcommand*{\SACLAY}{CEA-Saclay, Service de Physique Nucl\'eaire, 91191 Gif-sur-Yvette, France}
\affiliation{\SACLAY}
\newcommand*{\JLAB}{Thomas Jefferson National Accelerator Facility, Newport News, Virginia 23606}
\affiliation{\JLAB}
\newcommand*{\ITEP}{Institute of Theoretical and Experimental Physics, Moscow, 117259, Russia}
\affiliation{\ITEP}
\newcommand*{\ORSAY}{IPNO, Universit\'e Paris-Sud, CNRS/IN2P3, 91406 Orsay, France}
\affiliation{\ORSAY}
\newcommand*{\ANL}{Argonne National Laboratory, Argonne, Illinois 60439}
\affiliation{\ANL}
\newcommand*{\ASU}{Arizona State University, Tempe, Arizona 85287-1504}
\affiliation{\ASU}
\newcommand*{\UCLA}{University of California at Los Angeles, Los Angeles, California  90095-1547}
\affiliation{\UCLA}
\newcommand*{\CSU}{California State University, Dominguez Hills, Carson, CA 90747}
\affiliation{\CSU}
\newcommand*{\CMU}{Carnegie Mellon University, Pittsburgh, Pennsylvania 15213}
\affiliation{\CMU}
\newcommand*{\CUA}{Catholic University of America, Washington, D.C. 20064}
\affiliation{\CUA}
\newcommand*{\CNU}{Christopher Newport University, Newport News, Virginia 23606}
\affiliation{\CNU}
\newcommand*{\UCONN}{University of Connecticut, Storrs, Connecticut 06269}
\affiliation{\UCONN}
\newcommand*{\ECOSSEE}{Edinburgh University, Edinburgh EH9 3JZ, United Kingdom}
\affiliation{\ECOSSEE}
\newcommand*{\FU}{Fairfield University, Fairfield CT 06824}
\affiliation{\FU}
\newcommand*{\FIU}{Florida International University, Miami, Florida 33199}
\affiliation{\FIU}
\newcommand*{\FSU}{Florida State University, Tallahassee, Florida 32306}
\affiliation{\FSU}
\newcommand*{\GWU}{The George Washington University, Washington, DC 20052}
\affiliation{\GWU}
\newcommand*{\ECOSSEG}{University of Glasgow, Glasgow G12 8QQ, United Kingdom}
\affiliation{\ECOSSEG}
\newcommand*{\ISU}{Idaho State University, Pocatello, Idaho 83209}
\affiliation{\ISU}
\newcommand*{\INFNFR}{INFN, Laboratori Nazionali di Frascati, 00044 Frascati, Italy}
\affiliation{\INFNFR}
\newcommand*{\INFNGE}{INFN, Sezione di Genova, 16146 Genova, Italy}
\affiliation{\INFNGE}
\newcommand*{\JMU}{James Madison University, Harrisonburg, Virginia 22807}
\affiliation{\JMU}
\newcommand*{\KYUNGPOOK}{Kyungpook National University, Daegu 702-701, South Korea}
\affiliation{\KYUNGPOOK}
\newcommand*{\LPC}{LPC Clermont-Ferrand, Universit\'e Blaise Pascal, CNRS-IN2P3, 63177 Aubi\`ere, France}
\affiliation{\LPC}
\newcommand*{\LPSC}{LPSC, Universit\'e Joseph Fourier, CNRS/IN2P3, INPG, 38026 Grenoble, France}
\affiliation{\LPSC}
\newcommand*{\UMASS}{University of Massachusetts, Amherst, Massachusetts  01003}
\affiliation{\UMASS}
\newcommand*{\MOSCOW}{Moscow State University, General Nuclear Physics Institute, 119899 Moscow, Russia}
\affiliation{\MOSCOW}
\newcommand*{\UNH}{University of New Hampshire, Durham, New Hampshire 03824-3568}
\affiliation{\UNH}
\newcommand*{\NSU}{Norfolk State University, Norfolk, Virginia 23504}
\affiliation{\NSU}
\newcommand*{\OHIOU}{Ohio University, Athens, Ohio  45701}
\affiliation{\OHIOU}
\newcommand*{\ODU}{Old Dominion University, Norfolk, Virginia 23529}
\affiliation{\ODU}
\newcommand*{\RPI}{Rensselaer Polytechnic Institute, Troy, New York 12180-3590}
\affiliation{\RPI}
\newcommand*{\RICE}{Rice University, Houston, Texas 77005-1892}
\affiliation{\RICE}
\newcommand*{\URICH}{University of Richmond, Richmond, Virginia 23173}
\affiliation{\URICH}
\newcommand*{\SCAROLINA}{University of South Carolina, Columbia, South Carolina 29208}
\affiliation{\SCAROLINA}
\newcommand*{\UNIONC}{Union College, Schenectady, NY 12308}
\affiliation{\UNIONC}
\newcommand*{\VT}{Virginia Polytechnic Institute and State University, Blacksburg, Virginia   24061-0435}
\affiliation{\VT}
\newcommand*{\VIRGINIA}{University of Virginia, Charlottesville, Virginia 22901}
\affiliation{\VIRGINIA}
\newcommand*{\WM}{College of William and Mary, Williamsburg, Virginia 23187-8795}
\affiliation{\WM}
\newcommand*{\YEREVAN}{Yerevan Physics Institute, 375036 Yerevan, Armenia}
\affiliation{\YEREVAN}
\newcommand*{\NOWUNH}{University of New Hampshire, Durham, New Hampshire 03824-3568}
\newcommand*{\NOWUMASS}{University of Massachusetts, Amherst, Massachusetts  01003}
\newcommand*{\NOWMIT}{Massachusetts Institute of Technology, Cambridge, Massachusetts  02139-4307}
\newcommand*{\NOWECOSSEE}{Edinburgh University, Edinburgh EH9 3JZ, United Kingdom}
\newcommand*{\NOWGEISSEN}{Physikalisches Institut der Universitaet Giessen, 35392 Giessen, Germany}

\author {F.X.~Girod} 
\affiliation{\SACLAY}\affiliation{\JLAB}
\author {R.A.~Niyazov} 
\affiliation{\JLAB}
\affiliation{\RPI}
\author {H.~Avakian} 
\affiliation{\JLAB}
\author {J.~Ball}
\affiliation{\SACLAY}
\author {I.~Bedlinskiy} 
\affiliation{\ITEP}
\author {V.D.~Burkert} 
\affiliation{\JLAB}
\author {R.~De~Masi} 
\affiliation{\SACLAY}\affiliation{\ORSAY}
\author {L.~Elouadrhiri} 
\affiliation{\JLAB}
\author {M.~Gar\c con} 
\email[Corresponding author: ]{michel.garcon@cea.fr}
\affiliation{\SACLAY}
\author {M.~Guidal} 
\affiliation{\ORSAY}
% ...

\collaboration{The CLAS Collaboration}
     \noaffiliation
     
\date{\today}

\begin{abstract}
The beam spin asymmetries in the hard exclusive
electroproduction of photons on the proton ($\vec{e}p\to ep\gamma$)
were measured over a wide kinematic range and with high statistical accuracy.
These asymmetries result from the interference of the Bethe-Heitler
process and of deeply virtual Compton scattering. 
Over the whole kinematic
range ($x_B$ from 0.11 to 0.58, $Q^2$ from 1 to 4.8 GeV$^2$, $-t$
from 0.09 to 1.8 GeV$^2$), the azimuthal dependence of the asymmetries
is compatible with expectations from leading-twist dominance,
$A\simeq a\sin\phi/(1+c\cos\phi)$. This extensive set of data
can thus be used to constrain significantly the
generalized parton distributions of the nucleon in the valence quark
sector.

\end{abstract}

\pacs{12.40.Vv,13.40.Gp,13.60.Fz,13.60.Hb,13.60.-r,14.20.Dh,24.85.+p}

\maketitle

The structure of the nucleon, the lightest of all baryonic states,
has been studied in the past using two complementary approaches. 
Elastic electron
scattering measures form factors which reflect the spatial shape
of charge distributions~\cite{Hof61}, while deep inelastic scattering 
provides access to parton distribution functions that encode,
in a fast moving nucleon,
the momentum fraction carried by the constituents~\cite{Fri91}.
The 
formalism of Generalized Parton Distributions 
(GPDs)~\cite{Mue94,Ji97,Rad97} unifies these approaches and provides much
greater insight into nucleon structure~\cite{Ral01,Goe01}, through 
the coherence between states of different longitudinal momentum fractions, 
the correlation between transverse coordinates and longitudinal momentum 
of the partons~\cite{Bur00}, 
the distribution of forces exerted upon partons~\cite{Pol03}
(information inconceivable to obtain just a few years ago)
and the angular momentum carried by each type of parton~\cite{Ji97}.

Deeply virtual Compton scattering (DVCS) on the proton ($\gamma^*p\to\gamma p$),
in the Bjorken regime where the photon scattering occurs at the
quark level, is the process of choice to attain an experimental
determination of GPDs. Pioneering observations of DVCS
~\cite{Adl01,Air01,Ste01,Che03,Akt05,Che06,Air07a}, though of limited experimental
accuracy, are all compatible with a description of the observables
in terms of GPDs, both in the gluon and in the quark sector. 
Moreover, a recent precise experiment~\cite{Mun06}
gave good indications of the onset of scaling in this process 
at relatively modest values of the $\gamma^*$ virtuality. 

\begin{figure}
\hspace*{-1cm}\includegraphics[width=8cm]{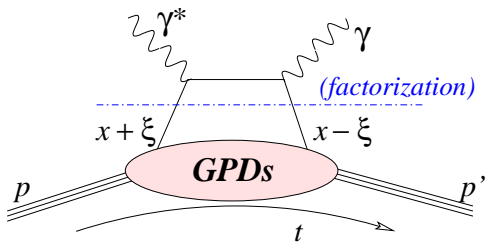}
\vspace*{-1cm}
 \caption{\label{fig:handbag}
 	   (Color online)
 	   Schematic representation of the leading-order handbag diagram 
 	   contribution to DVCS, where $x$ is the average longitudinal momentum 
 	   fraction of the active quark in the initial and final states 
 	   (measured in terms of the average hadron momentum $(p+p')/2$), 
 	   while $2\xi$ is their difference; it is related to the Bjorken scaling
 	   variable by $\xi \simeq x_B/(2-x_B)$. The squared four-momentum 
 	   transfer to the target is $t=(p'-p)^2$, and the squared four-momentum 
 	   of the virtual photon is $-Q^2$.
 	   }
\end{figure}

In this context, this work presents the first systematic and
precise exploration of a sensitive observable, the beam-spin
asymmetry of the reaction $\vec{e}p\to ep\gamma$. 
Neglecting a twist-3 DVCS term,
this asymmetry arises from the interference between the 
Bethe-Heitler (BH) and DVCS processes
(that is, where the photon is emitted by the electron or by the target
nucleon, respectively). At leading twist, it is primarily sensitive to
the imaginary part of the DVCS amplitude and thus to
a specific linear combination of the proton GPDs $H$, $\tilde{H}$ and $E$,
with arguments $x(=\pm\xi)$, $\xi$ and $t$.
Each proton GPD involves
a weighted sum over the quark flavors.
The beam-spin asymmetry is defined as
\begin{equation}
	A=\frac{d^4\vec{\sigma}-d^4\overleftarrow{\sigma}}
       		{d^4\vec{\sigma}+d^4\overleftarrow{\sigma}}\ ,
       	\label{eq:A_sigma}
\end{equation}
where the arrows correspond to beam helicity $+1$ and $-1$.
It depends on $Q^2$, $x_B$, $t$, defined in Fig.~\ref{fig:handbag},
and on the angle $\phi$ between the leptonic and hadronic planes.
Harmonic decompositions of the cross sections $d^4\sigma$, divided among 
contributions from
BH, DVCS and interference (INT) terms, have been proposed~\cite{Die97,Bel01}.
In the notation of Ref.~\cite{Bel01}, the cross sections, up to
some kinematic factors, can be expressed
in terms of the $\phi$-harmonics $c_n^{\text{S}}\cos n\phi$ and  
$s_n^{\text{S}}\sin n\phi$,
with $n$ from 0 to 3 and S = BH, INT or DVCS. 
At the twist-2 level,
which according to Ref.~\cite{Mun06} is largely dominant 
at least up to $|t| = 0.35$ GeV$^2$,
the numerator of Eq.~(\ref{eq:A_sigma})
gets a contribution from $s_1^{\text{INT}}$ only, 
while the denominator contains the coefficients 
$c_0^{\text{INT}}$, $c_1^{\text{INT}}$ and $c_0^{\text{DVCS}}$,
in addition to $c_n^{\text{BH}} (n=0,1,2)$ calculable in QED in terms of 
the proton elastic form factors. At leading twist, one obtains
\begin{equation}
	A = \frac{a\sin\phi}
       		{1+c\cos\phi+d\cos2\phi}\ ,
       	\label{eq:A_alpha}
\end{equation}
where the parameters $a$, $c$ and $d$ may be
expressed in terms of the above mentioned harmonic coefficients. 
The DVCS and INT harmonic coefficients may in turn be written
in terms of Compton form factors related to the corresponding GPD by
\begin{eqnarray}
	\Real {\cal H} &=& {\mathcal P}\int_{-1}^1\text{d}x
		\left[ {2x \over \xi^2-x^2 } \right] H(x,\xi,t)\ ,	
	\label{eq:ReCFF} \\
	\frac{1}{\pi}\Imag {\cal H} &=& H(\xi,\xi,t)-H(-\xi,\xi,t)\ ,
	\label{eq:ImCFF}	
\end{eqnarray}
up to corrections of order of the strong coupling constant,
with similar expressions for  
$\tilde{\cal H}$, ${\cal E}$ and $\tilde{\cal E}$.
The GPD $H$ 
yields 
the dominant contribution to the
harmonic coefficients considered above. Neglecting the
small contributions from the three other GPDs, one can express
the beam-spin asymmetry $A$ in terms of only 
$\Real {\cal H}$ and $\Imag {\cal H}$. 
Thus in this approximation, 
which is expected to hold for small values of $|t|$, the 
parameters $a$, $c$ and $d$ of Eq.~(\ref{eq:A_alpha})
are uniquely related to the imaginary and real parts
of the Compton form factor ${\cal H}$, yielding respectively
the GPD $H$ at points $x=\pm\xi$ and the principal value
integral of Eq.~(\ref{eq:ReCFF}). Going beyond this approximation
requires additional theoretical or experimental constraints on the
other GPDs.

The experiment took place in Hall B of Jefferson Lab, using
the CEBAF 5.77 GeV electron beam (with average polarization $P=0.794$),
a 2.5~cm-long liquid-hydrogen target and the CLAS spectrometer~\cite{Mec03}. 
The three final-state particles from the reaction $ep\to ep\gamma$
were detected. For this purpose, a new inner calorimeter (IC)
was added to the standard CLAS configuration,
55~cm downstream from the target,
in order to detect
1 to 5~GeV photons emitted between 4.5$^\circ$ and 15$^\circ$ 
with respect to the beam direction.
This calorimeter was built of 424 tapered lead-tungstate crystals,
16~cm long and with an average cross-sectional area of 2.1 cm$^2$, read out
with avalanche photodiodes and associated low-noise preamplifiers. 
The whole IC was operating at a stabilized temperature of 17$^\circ$C,
and monitored with laser light homogeneously distributed on all crystals.
The calorimeter was calibrated several times during the run using
the two-photon decay of neutral pions. Energy and angle resolutions
of 4.5\% and 4~mrad (for 1 GeV photons) were achieved.
In conjunction, a specifically-designed superconducting solenoid
was used to trap around the beam axis
the background originating from M{\o}ller electrons, 
while permitting detection of the recoil protons up to 60$^\circ$.

Events were selected if an electron had generated the trigger,
one and only one proton was identified and 
only one photon (above an energy threshold of 150 MeV) was detected in
either the IC or the standard CLAS calorimeter EC. Electrons were identified
through signals in the EC and in the \v{C}erenkov counters.
From time-of-flight information, track length and momentum, 
protons were unambiguously
distinguished from positive pions over the whole momentum range of interest.
All clusters detected in the IC were assumed to originate from photons,
while additional time-of-flight information was used in the EC to separate
photons from neutrons. For all three final-state particles, fiducial cuts 
were applied to exclude detector edges.
 
\begin{figure}
 	\includegraphics[width=\linewidth]{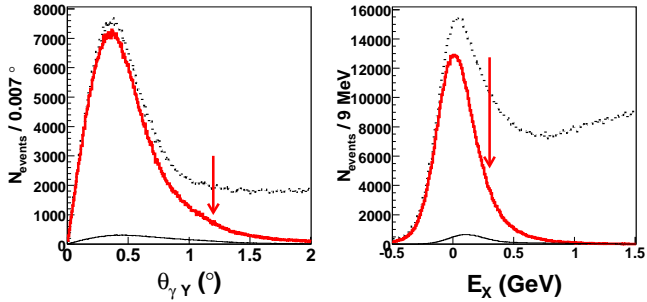}
 \caption{\label{fig:event_id}
 	   (Color online)
 	   Distributions in cone angle $\theta_{\gamma Y}$
 	   for the $ep\to epY$ reaction (left) and in
 	   missing energy $E_X$ for the $ep\to ep\gamma X$ reaction
 	   (right), before (black dotted curve) and after (red solid)
 	   all kinematic cuts discussed
 	   in the text but the one on the histogrammed variable,
 	   given by the location of the arrow.
 	   The thin solid black line represents the physical
 	   background, calculated from measured $ep\to ep\pi^0$ events.
 	   The distributions are integrated over all kinematic 
 	   variables and apply to the case where the photon is detected 
 	   in the IC.
 	 }
\end{figure}

Operating at a luminosity of 
$2\times10^{34}$~cm$^{-2}$s$^{-1}$ (a record for CLAS),
the accidental coincidences were negligibly small,
as well as the pile-up probability in the IC, except for the most 
forward photons below 6$^\circ$. 
Events considered here include the kinematic requirements~:
$Q^2>1$~GeV$^2$, $\gamma^*p$ invariant mass $W>2$~GeV and 
scattered electron energy $E'>0.8$~GeV. The mere selection of 
the three final-state particles results in the observation
of characteristic peaks in distributions of all kinematic variables
expressing the conservation of total four-momentum in the reaction
$ep\to ep\gamma$, as exemplified by the dotted curves in 
Fig.~\ref{fig:event_id}. Requiring in addition 
a missing transverse momentum smaller than 0.09 GeV,
an angle between the $\gamma^*p'$ and $\gamma p'$ planes smaller
than 1.5$^\circ$,
a photon detected within 1.2$^\circ$ of the direction
inferred from the detected electron and proton,
and a maximal missing energy $E_X$ of 0.3 GeV, 
results in clean peaks for the events of interest.
These kinematic cuts are to some extent redundant
(except for the background to be discussed below) and
are quoted here for the case where the emitted photon is
detected in the IC, that is for 92\% of the events. In the
case of photons detected in the EC, these cuts are about twice
as large because of the poorer resolution.

In spite of this selection, a contamination of events
originating from the $ep\to ep\pi^0$ reaction, followed by
the subsequent asymmetric decay of the neutral pion, is always 
possible. For these events, one
of the photons is not detected, because it is either below threshold 
or outside the calorimeters' acceptance. 
This
physical background is estimated using the number $N_{\pi^0}^{2\gamma}$ of
measured $ep\to ep\pi^0$ events, identified unambiguously when the two
photons are detected~\cite{DeM07}, and multiplying it by the ratio of acceptances
Acc$_{\pi^0}^{1\gamma}$/Acc$_{\pi^0}^{2\gamma}$, where the ``${1\gamma}$''
acceptance is to be understood with the photon satisfying all the
$ep\to ep\gamma$ event selection cuts. 
This ratio, which depends mostly on the
photon geometrical cuts and on the relevant resolutions, has been
calculated with the standard CLAS simulation package and a simplified
fast Monte-Carlo, the two results being used to evaluate the corresponding
systematic uncertainties. 
The background proportion $f$ varies between 1 and 25\% 
depending on the kinematic bin, 5\% in average.
The number of $ep\to ep\gamma$ events is
then, for each beam-helicity state and for each elementary bin
in the four kinematic variables (see below),
$\vec{N}
=\vec{N}_{ep\to ep\gamma X}
 - ($Acc$_{\pi^0}^{1\gamma}/$Acc$_{\pi^0}^{2\gamma})\vec{N}_{\pi^0}^{2\gamma}$,
and the asymmetry $A= (\vec{N}-\overleftarrow{N})/P(\vec{N}+\overleftarrow{N})$.
Finally, radiative corrections were applied~\cite{Gui07}.
These tend to increase the asymmetries very slightly.

\begin{figure}
 	\hspace*{-3mm}\includegraphics[width=88mm]{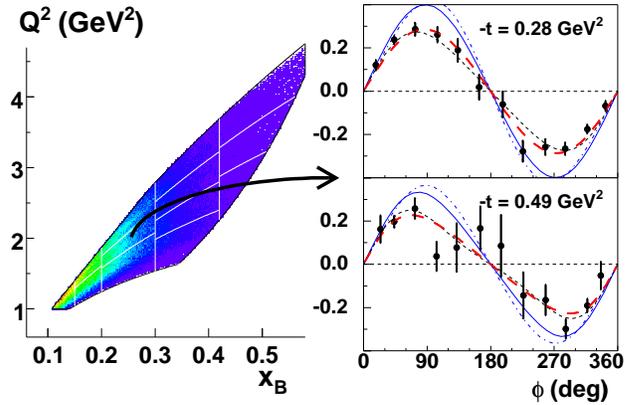}
 \caption{\label{fig:result1}
 	   (Color online)
 	   Left: kinematic coverage and binning in the ($x_B$, $Q^2$) space. 
 	   Right: $A(\phi)$ for 2 of the 62 ($x_B$, $Q^2$, $t$) bins,
 	       corresponding to
 	       $\langle x_B\rangle =0.249$, 
 	       $\langle Q^2\rangle =1.95$ GeV$^2$, 
		and two values of $\langle t\rangle$.
		The red long-dashed curves correspond to fits with
		Eq.~(\ref{eq:A_alpha}) (with $d=0$).
		The black dashed curves correspond
		to a Regge calculation~\cite{Lag07a}.
		The blue curves correspond to the GPD calculation
		described in the text,
		at twist-2 (solid) and twist-3 (dot-dashed) levels,
		with $H$ contribution only.
		}
\end{figure}

\begin{figure}
 	\hspace*{-3mm}\includegraphics[width=90mm]{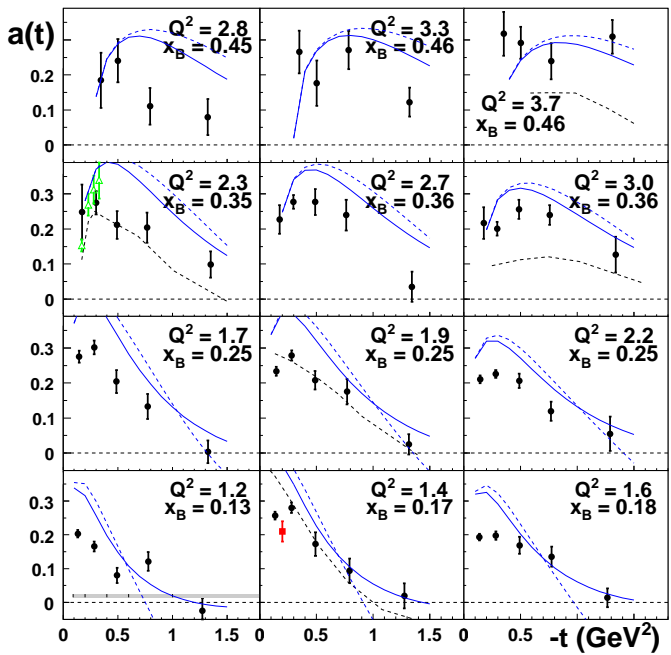}
 	\vspace*{-5mm}
 \caption{\label{fig:result2}
           (Color online)
 	   $a = A(90^\circ)$ as a function of $-t$.
	   Each individual plot corresponds to a bin in $(x_B, Q^2)$.
	   Systematic uncertainties and bin limits are illustrated 
	   by the grey band in the lower left plot.
 	   Black circles are from this work.
	   Previous results are from Ref.~\cite{Ste01} (red square) 
	   or extracted from cross section measurements~\cite{Mun06} 
	   (green triangles),
	   at similar - but not
 	   equal - values of $\langle x_B \rangle$ and $\langle Q^2 \rangle$.
	   See Fig.~\ref{fig:result1} caption for curve legend.
 	 }
\end{figure}

The data were divided into thirteen bins in the ($x_B$, $Q^2$) space
as per Fig.~\ref{fig:result1}, five bins in $-t$ 
(defined by the bin limits 0.09, 0.2, 0.4, 0.6, 1 and 1.8 GeV$^2$)
and twelve 30$^\circ$ bins in $\phi$.
Bin-size corrections were applied. 
Whether integrated in $t$ 
or in each $t$-bin (Fig.~\ref{fig:result1}),
the $\phi$-distributions were always found to be compatible with 
Eq.~(\ref{eq:A_alpha}) with $d=0$.
The parameter $d$ is expected to be smaller than 0.05 over our
kinematic range, and indeed
was found compatible with zero, within statistical accuracy,
when including it in the fit. 
The deviation from a pure sine function
as $|t|$ increases is seen in all $(x_B,Q^2)$ bins
and results in the parameter $c$ becoming negative~\cite{Gir06}.
The parameter $a$ is
the best estimate of $A(90^\circ)$ and is represented
in Fig.~\ref{fig:result2}.
Point-to-point systematic uncertainties arise mostly from 
the background subtraction:
$\Delta A_b=(A-A_{\pi^0})\Delta f/(1-f)$, where the relative error
on $f$ is conservatively estimated to be 30\% and $A_{\pi^0}$ is the
asymmetry for the reaction $ep\to ep\pi^0$, ranging  between
0.04 and 0.11 at 90$^\circ$~\cite{DeM07}.
The sensitivity of the results to the event selection cuts was
studied as well. 
From these two sources of information, 
the systematic uncertainty on $a$ was inferred 
to be 0.010, independent of $x_B$, $Q^2$ and $t$.
An overall normalization uncertainty arises from the uncertainty
in the beam polarization (3.5\%).
Additional details on the experiment and on the
data analysis may be found in Ref.~\cite{Gir06}.

The wide kinematic coverage of the present data is important for
global analyses of $ep\to ep\gamma$ observables and for a model-independent extraction 
of DVCS amplitudes. The beam-spin asymmetries are especially, but not uniquely, 
sensitive to the GPD $H$. 
When combined with other observables more sensitive
to $\tilde{H}$ and $E$, as well as
with unpolarized cross sections, it will be possible
to obtain the real and imaginary parts of the Compton form factors
of all GPDs, as defined in Eqs.~(\ref{eq:ReCFF}) and~(\ref{eq:ImCFF}).
Additional theoretical work is also required, to clarify how
power-suppressed contributions not included in Ref.~\cite{Bel01} would
affect the relations between observables and GPDs~\cite{Mue07}.
Presently,
GPDs may be calculated using theoretical models based on 
constituent quarks, on a chiral quark-soliton description
of the nucleon, on light-cone or other frameworks.
The first moments of GPDs are being calculated using
lattice QCD techniques. But none of these calculations are developed
to the point of making the link to DVCS observables. Alternatively, 
constrained parameterizations 
have been used to make predictions of DVCS beam-spin asymmetries.
Following Refs.~\cite{Rad01,Gui05}, 
such a parameterization of the GPD $H$ may be 
\begin{eqnarray}
	H\!\! &=&\!\! \sum_q e_q^2 \left\{\int_{-1}^{+1}\text{d}\beta
	                 \int_{-1+|\beta|}^{1-|\beta|}\text{d}\alpha\,
	                 \delta(x-\beta-\xi\alpha)
                  \mathfrak{h}^q(\beta,\alpha,t)\right. \nonumber\\
                 &&\ \ \ \ \ \ \ \ \ 
    		+\left.\theta\left( 1-\frac{x^2}{\xi^2}\right) 
    		D^q\left(\frac{x}{\xi},t\right)\right\}\ ,
\label{eq:H_DD}
\end{eqnarray} 
\begin{equation}
\text{with}\ \ \ \ \
\mathfrak{h}^q(\beta,\alpha,t)=q(\beta)\pi_b(\beta,\alpha)e^{-\alpha'_1(1-\beta)t}\ ,\ \
\label{eq:DD}
\end{equation}
where $e_q$ and $q(\beta)$ are the electric charge and
unpolarized parton distribution for quark flavor $q$, 
$\pi_b$ a profile function~\cite{Rad01} and $\alpha'_1$ is a Regge slope
adjusted to recover the proton form factor $F_1$ from the first moment of the
GPD. 
Eq.~(\ref{eq:DD}) extends the ansatz of Ref.~\cite{Gui05} for the
$t$ dependance to non-zero values of $\xi$.
The $D$ term in Eq.~(\ref{eq:H_DD}) is calculated 
within a quark-soliton chiral model~\cite{Goe01}. 
Using predetermined
parameters, the  calculations of beam-spin asymmetries
yield the
solid and dot-dashed curves in Figs.~\ref{fig:result1} and~\ref{fig:result2}, 
without and with a twist-3 term calculated in the Wandzura-Wilczek 
approximation~\cite{Goe01}. 
The predictions overestimate the asymmetries at low $|t|$, 
especially for small values of $x_B$ and/or $Q^2$.
Variations of the parameter $b$ entering the
profile function $\pi_b$ do not resolve this problem, which may indicate
that double distributions are not flexible enough to reproduce this
behaviour. 

Alternatively, description
of the process in terms of meson (or more generally Regge
trajectory) exchanges has been attempted~\cite{Lag07a,Szc07}.
DVCS may be viewed as $\rho$ production followed by
$\rho$-$\gamma$ coupling in vacuum or in the nucleon field. 
In addition to pole contributions
in the $t$ channel~\cite{Can03}, the box diagram
that takes into account  $\rho$-nucleon intermediate states
has been evaluated~\cite{Lag07a}. 
This calculation, represented by the dashed curves in 
Figs.~\ref{fig:result1} and~\ref{fig:result2}, is in fair
agreement with our results up to $Q^2=2.3$ GeV$^2$. 
The significance 
of this dual description (Regge vs. handbag) remains to be fully investigated.

In summary, the most extensive set of DVCS data to date has been obtained
with the CLAS spectrometer, augmented with specially designed 
small-angle photon calorimeter and solenoid.
Beam-spin asymmetries were extracted in the valence quark region,
as a function of all variables describing the reaction. 
Present parameterizations of 
GPDs describe reasonably well, but not perfectly, the main features
of the data.
The measured kinematic
dependences will put stringent constraints on any DVCS model,
and in particular on the generalized parton distributions in the nucleon.

\begin{acknowledgments}
We would like to thank B. Hervieu (DAPNIA-Saclay), Ph. Rosier (IPN-Orsay)
and their collaborators
for the skillful engineering of the new equipment. 
We acknowledge the outstanding efforts of the staff of the 
Accelerator and Physics Divisions at Jefferson Lab 
that made this experiment possible.
We also acknowledge useful discussions with D. M\"{u}ller.
This work was supported in part by 
the French Centre National de la Recherche Scientifique 
and Commissariat \`{a} l'Energie Atomique,
the U.S. Department of Energy 
and National Science Foundation,
the Italian Istituto Nazionale di Fisica Nucleare,
the Korean Science and Engineering Foundation, 
the U.K. Engineering and Physical Science Research Council.
The Jefferson Science Associates (JSA) operates the 
Thomas Jefferson National Accelerator Facility for the United States 
Department of Energy under contract DE-AC05-06OR23177. 

\end{acknowledgments}

\bibliography{biblio}

\end{document}